\documentclass[prb,twocolumn]{revtex4-1} 

\usepackage[breaklinks,colorlinks,citecolor=blue]{hyperref}

\usepackage{graphicx}

\usepackage[utf8]{inputenc}

\usepackage{amsfonts}
\usepackage{amssymb}

\usepackage{tikz}

\newcommand{\be}{\begin{equation}}
\newcommand{\ee}{\end{equation}}
\newcommand{\bea}{\begin{eqnarray}}
\newcommand{\eea}{\end{eqnarray}}

\renewcommand{\doi}[1]{{\sc doi:} \href{http://dx.doi.org/#1}{\detokenize{#1}}}

\usepackage{etoolbox}
\newtoggle{anonymizedVersion}
\settoggle{anonymizedVersion}{true}
\settoggle{anonymizedVersion}{false}

\newcommand{\makeAnon}[2]{\iftoggle{anonymizedVersion}{#1}{#2}}

\begin{document}

\title{``It’s getting away from us!'' --- Black Hole Horizons and Relative Speed}

\author{\makeAnon{Name anonymized}{M P{\"o}ssel}}
\address{\makeAnon{Address anonymized}{Haus der Astronomie and Max Planck Institute for Astronomy, K{\"o}nigstuhl 17, 69124 Heidelberg, Germany}}
\email{\makeAnon{nemo@anonymized.org}{poessel@hda-hd.de}}

\begin{abstract}
Black holes hold considerable fascination for the general public and students alike, and are commonly included in general-science courses on astronomy or modern physics. But teaching about the basics of black holes poses a considerable challenge: Any rigorous description requires concepts and techniques from general relativity, Einstein’s theory of geometry and gravitation. 

And any half-way rigorous introduction to that theory, including the required mathematical tools, is significantly beyond the level of general-science courses. Inevitably, accounts of relativistic physics at the introductory undergraduate level make use of analogies, approximations and simplified models to teach about topics like black holes, gravitational waves, gravitational lensing, or cosmology. 

The purpose of this article is to given an account of one particular set of analogies for teaching about black holes, all of which are based on modelling the motions of observers in the vicinity of the black hole and rely on the concept of (relative) speed to describe properties of the black hole. While most elements of what I am about to describe can be found in the existing literature, I am not aware of any text that attempts to pull them together into a unified picture at a suitable level of presentation for undergraduate-level teaching; that is the goal of the present text.\footnote{The following article has been accepted by {\em The Physics Teacher}. After it is published, it will be found at \url{https://pubs.aip.org/aapt/pte}}
\end{abstract}
\maketitle 

\section{Black holes and their horizons}

The effect of mass on space and time is at the heart of general relativity, and black holes are an extreme example: Objects beyond a certain level of compactness, that is, objects where a sufficiently high mass is confined within a region of space with a sufficiently small diameter, create an isolated region with a boundary that is called a horizon. Einstein’s theory predicts that in this kind of situation, no physical object entering from the outside will ever be able to leave the horizon-bounded region. Neither can light from inside the horizon reach any observer on the outside. Those self-isolated regions are what we call black holes. 

The existence of black holes has interesting astrophysical consequences, from active galactic nuclei to gravitational waves to the recent images of “black hole shadows” by the Event Horizon Telescope collaboration. Popular- and introductory-level descriptions of the fundamentals, the astrophysical context, and the historical aspects can be found in numerous books and articles.\cite{PopularBH}  Here, we will instead focus on one particular aspect of the basics: understanding the horizon and the redshift effects associated with it.

\section{The river model of black holes}
The ``river model of black holes'' illustrates the simplest kind of black hole, which is spherically symmetric and static. The model was first introduced into modern astronomy education in 2008,\cite{Hamilton2008}  but the underlying mathematics goes back to the 1920s.\cite{Painleve1921,Gullstrand1922}

\begin{figure*}
\includegraphics[width=0.75\textwidth]{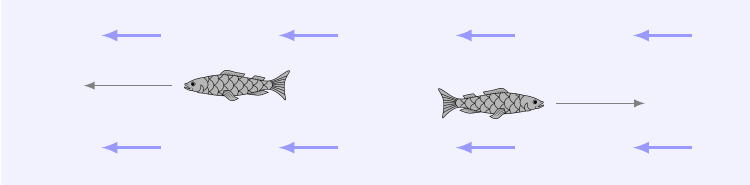}
\caption{Cross section of a river with fish swimming at the speed of fish. Blue arrows indicate flow speed, grey arrows fish speed relative to the surrounding water}
\label{fig:river-no}
\end{figure*}

The basic idea is as follows: Consider a river, as in Fig.~\ref{fig:river-no}. As an observer on the river bank, you can see not only the flowing water, but also a peculiar species of fish. Those fish swim at constant speed $c$, which we will call the ``speed of fish.'' In a river flowing at constant speed $v$, you will see fish swimming upstream making progress at the speed $c-v$ relative to the river bank, while fish that go with the flow move at $c+v$.

\begin{figure*}
\includegraphics[width=0.75\textwidth]{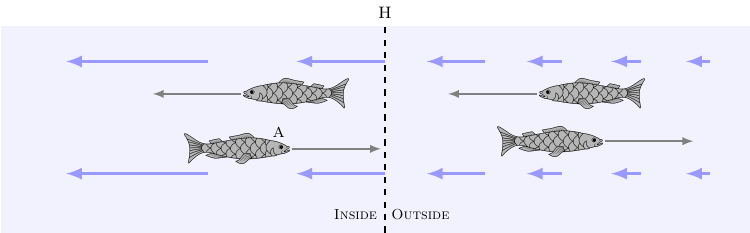}
\caption{River whose speed of flow increases from right to  left, surpassing the speed-of-fish at the river horizon H}
\label{fig:river-horizon}
\end{figure*}

Next, consider a river whose flow speed is increasing continually from right to left, as pictured in Fig.~\ref{fig:river-horizon} This could be a river which gets narrower towards the left, forcing water speed to increase so as to ensure a steady flow. At the point H, marked here on the river bank, the flow speed surpasses the speed of fish, which creates a curious situation: As external observers, we will never see a fish from the region to the left of H (re-)enter the region to the right of H! In order to do so, that fish would need to swim faster than the speed of fish, which is impossible in our model. 

H is a river horizon: Fish can move freely to the left or to the right if they are “outside of H,” that is, to the right of H. But once they have passed H, external observers will only ever see them move to the left --- even fish A in Fig.~\ref{fig:river-horizon}.

This is a helpful analogy for what happens near a black hole. Students can think of ``space itself'' flowing into the black hole at a speed that increases as we approach the black hole. Since light will only ever move at the speed of light, once we have reached the point where space is falling, or flowing, into the black hole at faster-than-light speeds, light from such a region will not be able to make its way outward. 

The boundary where space is falling at exactly the speed of light, for us as external observers, marks the point of no return: the black hole’s (event) horizon. Outside of the horizon, light can move towards or away from the center of symmetry, the ``center of the black hole'' for short. Inside the horizon, light can only move ever further inward. 

Another interesting analogy with the river model: Locally, fish will not be able to tell when they have crossed the horizon. The water around them looks just the same to the right of the horizon as to the left. Analogously, in classical general relativity, an observer in free fall will not feel any local effects of gravity: Such an observer will feel weightless, and will see light and matter in their immediate (infinitesimal) vicinity behave as if there were no gravity at all. This is known as {\em Einstein’s equivalence principle}. Specifically, the falling observer will note nothing unusual in their local measurements when passing the horizon.

\section{From unusual coordinates to the need for (relativistic, relative) speed}

The model is readily generalized to three-dimensional space, with space “flowing” towards the center of the black hole radially, from all directions. In 3D, the black hole horizon is a spherical surface. 

If this is your first encounter with the river model, you might well be skeptical. The model does provide for a neat analogy. But is it more than just a superficial illustration? As it turns out, the model is much more. Behind the simplified picture is a rigorous mathematical description, based on a particular set of alternative coordinates for the famous Schwarzschild solution for the spacetime geometry around a simple black hole.\cite{Hamilton2008,Painleve1921,Gullstrand1922}  And the model is not limited to the simplest, spherically-symmetric case: it is possible to generalize the river model to describe rotating black holes,\cite{Hamilton2008} and even black holes in an expanding universe.\cite{Braeck2013} 

\section{An ether in disguise?}

The genesis of special relativity is often told as Einstein doing away with the concept of absolute motion, and specifically with the concept of an ``ether'' as a preferred state of rest. Does the river model, with its talk of space ``falling'' or ``flowing,'' re-introduce the ether, through a backdoor? 

There are several aspects to disentangle here. The original ether provided a background structure that was present regardless of the specific physical situation. It was linked to the idea that, by making local measurements of the speed of light, you should be able to determine the speed of your own reference frame with respect to the ether. The possibility to do so was encoded in the basic laws of physics, without any references to specific objects residing in space. 

The ``river'' situation is different: It is linked to the presence of the black hole, that is, with a specific configuration of mass in space, not with the underlying laws of general relativity. This is true for all spacetime changes in general relativity that are caused by sources of gravity, which makes for a fundamental difference between those contingent structures and the original ether.\cite{StachelComment}

So what, then, is meant by ``falling'' or ``flowing'' space in the descriptions above? Mathematically, these statements correspond to a specific choice of coordinates for describing space and time in this particular situation. Physically, the choice made for the river model can be understood as filling spacetime with the trajectories of infinitely many test particles, which start out at rest infinitely far away from the black hole and fall straight towards the black hole (which means that they have zero angular momentum). 

Idealized clocks carried along by those particles, and synchronized so as to preserve the spherical symmetry, are then used to define a global time coordinate: At each point in space, and at each moment in time, there will be one of this particular family of infalling test particles, and our choice of coordinates means that at each such spacetime point, we are choosing to describe local physics using the idealized clock attached to the infalling test particle in question. Taken together, information about all these test particles tells us everything we need to know about the black hole affecting motion in its surrounding space. 

In general relativity, separating spacetime into space and time is always linked to a specific choice of time coordinate --- and the laws of general relativity by themselves do not privilege one choice above another. Space as defined by this particular choice of time coordinate ``flows towards the black hole.'' Describing spacetime in this way does not amount to establishing an absolute, obligatory pattern of motion. Other types of motion, with different initial velocities, and non-zero angular momentum, remain possible, as do infinitely many other choices of time coordinate that do not amount to ``radially flowing space.'' 

But this particular coordinate choice, and its associated pattern of motion, is useful in that it provides us with an easy-to-understand way of identifying the black hole’s event horizon. That horizon’s existence and location {\em are} physical properties of the underlying spacetime. Finding a horizon would be the outcome regardless of choice of coordinates in the spacetime region in question. 

\section{Relativistic relative speed}
Maybe you are still skeptical. I have just explained that the ``flow of space'' is a consequence of a certain choice of time coordinate, and you might have heard that, in general relativity, one should not blindly trust coordinate-based statements. Isn’t there some physical, coordinate-independent way of deciding whether what is happening at the black hole horizon involves motion, a ``moving away'' from us? The answer is that there is, even though it’s more complicated than in classical physics.

Even in special relativity, when you want to describe two objects’ speed relative to each other, you need to do more than merely point out which objects you are talking about. After all, the velocity of objects can change, and there is no universal notion of simultaneity. What you can define is the speed of an object at specified time $t_1$ relative to a second object at specified time $t_2$. You can always find an inertial system $I_1$ --- one of the preferred reference systems in special relativity --- that is at rest relative to the first object at time $t_1$, and another inertial system $I_2$ that is at rest relative to the second object at time $t_2$. The (constant) speed of $I_1$ as measured from $I_2$ is the relative speed you are looking for. With this procedure, in the language of spacetime, you have specified two events, one each on each object’s worldline, in order to define relative speed.

In general relativity, the situation is even more complicated, and the details are beyond what you can describe without invoking the formalism of general relativity (but then, so is the geometry of black holes). There, if you want to compare two velocities, you need not only to specify two events, but a ``spacetime path'' between those events --- and for different paths, you will in general obtain different results! 

Fortunately, an important type of situation has ``built-in'' paths that we can use for comparison: Whenever we send a light signal from one observer to another, and want to compare the motion of the first observer at the moment of light emission relative to the second observer at the moment of the reception of the light, the light’s own trajectory (``light-like geodesic'') provides a natural path for the comparison.\cite{Narlikar1994} 

If you were to perform this relativistic relative speed calculation for our black hole spacetime, you would find, and now in a properly coordinate-independent way, that yes, the event horizon is the limiting point where the speed relative to an external observer of an infalling object, and remarkably even the speed of a spaceship that tries to slow down its descent with all engines firing, approaches the speed of light. Horizons are where things are getting away from us at the speed of light!

\section{Redshifts near a black hole as Doppler shifts}

Viewing space as moving (``flowing'') into the black hole, and applying the relativistic definition of relative speed, has another advantage. Near a black hole, there are characteristic phenomena associated with light propagation and with wavelength shifts (equivalently: frequency shifts) for light. When two observers hover at different constant distances above the black hole horizon, one below the other, the higher-up observer will observe light emitted by the lower-down observer as redshifted, and the lower observer will see light received from their higher-up colleague as blueshifted. 

This is commonly referred to as the gravitational redshift, and it can be calculated in a straightforward manner from Einstein’s equivalence principle.\cite{Poessel2021}  Place the lower-down observer closer and closer to the black hole horizon, and the redshift measured by the higher-up observer increases beyond all bounds. For a light source at the horizon itself, the gravitational redshift becomes infinite, which is another way of saying that no light reaches the higher-up observer at all.

In the river model, it is natural to interpret those wavelength shifts as a direct consequence of motion, in other words: as Doppler shifts. Where the horizon is a region ``getting away from us at the speed of light,'' it is natural to interpret the infinite wavelength shift for light signals from that region as the associated Doppler shift. The relativistic Doppler formula for an object moving directly away from us at a speed $v$ is
\be
z\equiv \frac{\lambda-\lambda_0}{\lambda_0}=\sqrt{\frac{c+v}{c-v}}-1
\ee
where $\lambda_0$ is the wavelength at which the light is emitted, $\lambda$ the wavelength at which it is received. As $v$ approaches the speed of light $c$, this expression diverges. At least for signals sent from the horizon, the redshift is exactly what we would expect for a region ``getting away from us at the speed of light.''

\section{Doppler effects in a static situation?}
Here you might again be skeptical. It’s one thing to talk about the unusual infinite redshift at the horizon as a Doppler effect, but how can motion even begin to explain the redshift, or blueshift, between two hovering observers, each at constant distance from the black hole, one below the other? Doesn’t ``constant distance from one each other'' imply that two such observers are at rest relative to each other, and that we shouldn’t expect a Doppler shift at all? 

And even if there were relative motion involved, with an associated Doppler shift, shouldn’t we expect it to be either one or the other --- a blueshift for relative motion towards each other, or else a redshift, associated with motion away from each other? At first glance, it might seem as if this configuration is beyond any Doppler interpretation: How can a Doppler shift result in both a blueshift for a signal from the higher-up to the lower observer, and a redshift when the signal goes the other way? Shouldn’t relative motion be either one or the other?

The answer turns out to be that, on the contrary, those seemingly contradictory properties for Doppler shifts are generically true whenever accelerated motion is involved --- even in special relativity.\cite{Poessel2024}  A key part of the effect can even be seen in classical, non-relativistic physics!

\section{Distance measurements as radar measurements}
By definition, the velocity of an accelerated observer changes over time. A statement like ``there is a Doppler shift for signals passing between these observers'' unavoidably involves delay: For the Doppler shift, we need to compare the state of motion of the source at the time of emission with that of the receiver at the time of reception. And unlike in classical physics, in relativity, even a statement about objects being at rest relative to each other is not a comment on some “objectively instantaneous” property of the situation. The standard coordinates in special relativity are based on radar measurements:  An inertial observer (the preferred kind of observer in that theory) measures distances by sending a light signal to another object at time $t_e$, receiving the reflected light signal at time $t_r$, and deducing the distance to be
\be
d=\frac{c}{2}(t_r-t_e ).     
\ee           
This is directly tied in with the role of the speed of light $c$ in special relativity and with Einstein’s definition of simultaneity. Even a simple statement like ``the distance between myself and object A is constant'' involves measurements at different points in time. Applying this knowledge to accelerated systems yields the above-mentioned, somewhat counterintuitive results for relative motion and the Doppler effect.

\section{Acceleration and unusual Doppler shifts: a simple example}

\begin{figure}
\includegraphics[width=0.9\linewidth]{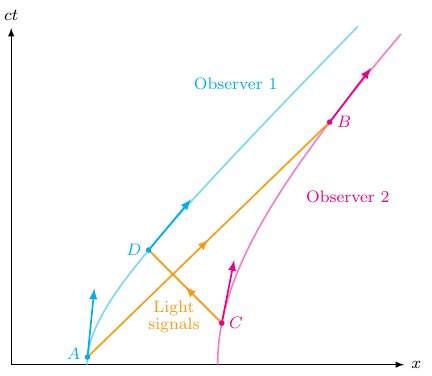}
\caption{Accelerated observers exchanging light signals. Coordinates are chosen so that light worldlines (yellow) are exactly diagonal. At the event $A$, the first observer sends out a light signal, which is received by the second observer at the event $B$. At event $C$, the second observer sends a light signal towards the first, which the first receives at event $D$. Also shown are the tangents to the observers’ world lines at these four events, which indicate the observers’ velocities at each event. The comparison between these velocities shows that observer 1 at $A$ is moving away from observer 2 at $B$, while observer 2 at $C$ is moving towards 1 at $D$. Evidently, in an accelerated situation, it is possible for observer 1 to perceive light from observer 2 as blue-shifted, while observer 2 perceives light from observer 1 as redshifted.}
\label{fig:acc-example-radar}
\end{figure}

As an example, consider Fig.~\ref{fig:acc-example-radar}, a spacetime diagram with the worldlines of two accelerated observers and of two light signals propagating between them. 

At the event $A$, observer 1 emits a light signal, which observer 2 receives at event $B$. Observer 1’s velocity at $A$ is slower than observer 2’s at $B$, as can be read off the diagram: The tangent to observer 1’s worldline at $A$ is steeper than the tangent to observer 2’s worldline at $B$. In consequence, observer 1 at $A$ and observer 2 at $B$ are moving away from each other, and observer 2 receives the light signal as redshifted. A similar line of argument shows that the light signal sent out by observer 2 at $C$ is received as blue-shifted by observer 1 at $D$. (This is the part of the argument that holds even in classical physics!) 

\begin{figure}
\includegraphics[width=0.9\linewidth]{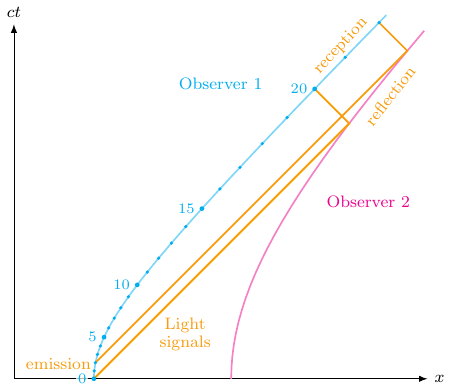}
\caption{Accelerated observers exchanging light signals. Markings on the worldline of observer 1 and the associated numbers indicate observer 1’s proper time: the time measured by an ideal clock moving along that worldline.  Coordinates are chosen so that light worldlines (yellow) are exactly diagonal. Observer 1 emits two light signals towards observer 2, which observer 2 reflects and sends back. The proper time markings on observer 1’s world line show that the time interval between emission and reflection is the same for both signals. Observer 1 concludes that the radar distance between herself and observer 2 is constant.}
\label{fig:acc-example-constant}
\end{figure}

And yet, as Fig.~\ref{fig:acc-example-constant} illustrates for two radar measurements by observer 1, the radar distances between the two observers are constant: radar distance is defined for a light signal emitted at time $t_e$  and the reflected signal received at time $t_r$ as $d=\frac12 [ct_r-ct_e ]$. As this is a radar measurement by observer 1, $t_e$ and $t_r$ in the formula need to refer to a clock that is at rest at observer 1’s location, moving along with observer 1.

In general relativity, time measured along an observer’s world line by a co-moving clock in this way is called the world line’s proper time. In Fig.~\ref{fig:acc-example-constant}, equal-time intervals of proper time $c\tau$ are marked on observer 1’s worldline, and from the time stamps for the emission and reception of the two light signals shown in that figure, we can tell that the radar distance remains the same: $\frac12 [20-0]=\frac12 [22-2]=10$. 

Calculations confirm this for the fully relativistic description: Contrary to most people’s (and most physicists’?) intuition, for accelerated objects, constant (radar) distance does not preclude relative motion, and the associated Doppler effects can have opposite signs. 

Thus, there is no contradiction. If we are dealing with accelerated observers, it is quite possible to measure a blueshift or redshift from another observer while at the same time finding radar distances to be constant. It is also possible for there to be a redshift-blueshift asymmetry, with observer 1 measuring observer 2’s signals to be blueshifted, while observer 2 measures observer 1’s signals as redshifted. 

In conclusion: In terms of mathematical formulae, a Doppler shift and a gravitational redshift might look quite different.\cite{Simionato2021}  But on closer inspection, and making use of the relativistic definition of relative motion, the classical gravitational redshift near a black hole (or any other mass, for that matter) can equivalently be described as a Doppler shift. The classical and the Doppler description are merely two different ways of describing the same underlying phenomenon.

\section{Summary}
The river model of black holes provides a helpful picture that can help students to understand the basic properties of black holes. As we have seen, that picture can be extended in a consistent way to key properties of black hole spacetimes: The view of black hole horizons (and, incidentally, also cosmological horizons\cite{Poessel2020}) as ``regions of space moving away from us at the speed of light'' provides an intuitive understanding of why no light from behind a horizon can reach us. 

That moving-away can be expressed in terms of a relativistic relative speed, and the associated wavelength shifts of light understood as Doppler shifts. What might at first seem like internal contradictions of the interpretation, namely the constant distances for static observers, and the different signs of the Doppler shifts involved (specifically, redshifts and blueshifts in the same situation) can be explained as well: Both are general properties of accelerated observers, present even in special relativity. As an additional mark of consistency, the Doppler interpretation directly implies infinite redshift at the horizon.

\section*{Acknowledgements}

I would like to thank Thomas M{\"u}ller and the anonymous reviewers for helpful feedback.

\end{document}